\documentclass[pra,twocolumn,floatfix,superscriptaddress,showpacs,eqsecnum,nofootinbib]{revtex4}

\usepackage{graphicx}
\usepackage{color} 
\usepackage{amsmath} 
\usepackage{amssymb}
\usepackage[colorlinks=false]{hyperref}
\usepackage{epstopdf}
\usepackage[normalem]{ulem}
\usepackage{subfigure}

\newcommand{\ket}[1]{\left|#1\right\rangle}

\newcommand{\braket}[2]{\left\langle#1\vert#2\right\rangle}

\newcommand{\mbf}[1]{\mathbf{#1}}
\newcommand{\ud}{\mathrm{d}}
\DeclareMathOperator{\Tr}{Tr}

\begin{document}
\title{High Fidelity Quantum Gates in the Presence of Dispersion}
\date{\today}
\author{B. Khani}
\affiliation{Institute for Quantum Computing and Department of Physics and Astronomy, University of Waterloo, Waterloo, Ontario, Canada N2L 3G1}
\author{S. T. Merkel}
\altaffiliation[Present address: ]{IBM T.J. Watson Research Center, Yorktown Heights, New York 10598, USA}
\affiliation{Institute for Quantum Computing and Department of Physics and Astronomy, University of Waterloo, Waterloo, Ontario, Canada N2L 3G1}
\affiliation{Theoretical Physics, Saarland University, 66123 Saarbr\"ucken, Germany}
\author{F. Motzoi}
\affiliation{Institute for Quantum Computing and Department of Physics and Astronomy, University of Waterloo, Waterloo, Ontario, Canada N2L 3G1}
\author{Jay M. Gambetta}
\altaffiliation[Present address: ]{IBM T.J. Watson Research Center, Yorktown Heights, New York 10598, USA}
\affiliation{Institute for Quantum Computing and Department of Applied Mathematics, University of Waterloo, Waterloo, Ontario, Canada N2L 3G1}
\author{F. K. Wilhelm}
\email[Email: ]{fwm@lusi.uni-sb.de}
\affiliation{Institute for Quantum Computing and Department of Physics and Astronomy, University of Waterloo, Waterloo, Ontario, Canada N2L 3G1}
\affiliation{Theoretical Physics, Saarland University, 66123 Saarbr\"ucken, Germany}

\begin{abstract}
We numerically demonstrate the control of motional degrees of freedom of an ensemble of neutral atoms in an optical lattice with a shallow trapping potential. Taking into account the range of quasimomenta across different Brillouin zones results in an ensemble whose members effectively have inhomogeneous control fields as well as spectrally distinct control Hamiltonians. We present an ensemble-averaged optimal control technique that yields high fidelity control pulses, irrespective of quasimomentum, with average fidelities above 98\%. The resulting controls show a broadband spectrum with gate times in the order of several free oscillations to optimize gates with up to 13.2\% dispersion in the energies from the band structure. This can be seen as a model system for the prospects of robust quantum control. This result explores the limits of discretizing a continuous ensemble for control theory.
\end{abstract}
\maketitle
\section{Introduction}

Optical lattices \cite{Pethick08} are a suitable platform for the simulation of strongly correlated many-body and other condensed matter systems \cite{Lewenstein07,BlochDalibard08,Bloch08}.  These simulations can in turn be used to compute otherwise intractable problems such as antiferromagnetic transitions \cite{Simon11}, and conductor to insulator transitions \cite{Jaksch99,Greiner02}, to name a few.   Optical lattices are also a promising candidate system for quantum information processing \cite{Deutsch00}. In the simplest case, a one-dimensional optical lattice is created by two counter-propagating beams of coherent light which create a standing wave.  This generates a trapping potential in which sufficiently cold atoms can be bound and trapped in a one-dimensional periodic potential but can move freely in the other two spatial dimensions.  It also provides an opportunity to investigate a difficult control problem in state evolution, particularly from the perspective of quantum computation with qubits formed by the vibrational states of atoms in the lattice, so-called external qubits \cite{Deutsch98,Chen06,Maneshi08,Schneider11}.
 
\begin{figure}[b!]
\centering
\includegraphics[width=1.0\columnwidth]{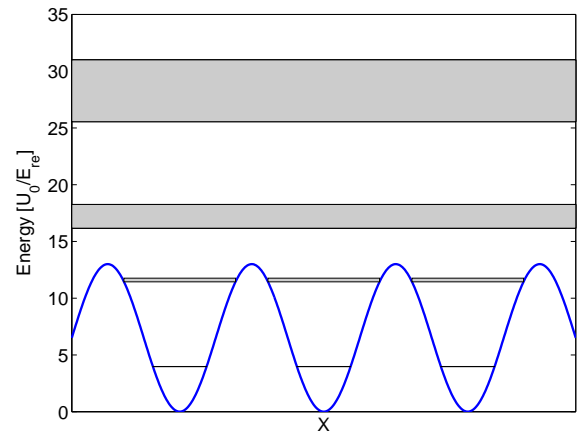}
\caption
{(color online) The first four energy bands (highlighted in grey) for a 1-D optical lattice are shown in the localized basis for potential depth $r=13$ and compared to the sinusoidal trapping potential. Refer to Fig. \ref{fig:energy}.}
\label{fig:localenergy}
\end{figure}

One of the strengths of optical lattices is that there is a great deal of freedom for control and manipulation, however, in practice finding the control sequences to perform a desired task can be difficult.  The control task we examine in this document is to perform a single-qubit gate between states in the lowest two Bloch bands.  The eigenstates of a periodic potential have a band structure, with states within the bands distinguished by a quasimomentum, see Fig.~\ref{fig:localenergy}.  Our target gates are naturally vertical, i.e., they connect states with equal quasimomenta. 
The controls, phase and intensity modulations of the control laser, both do not affect the discrete displacement symmetry hence they do not change the quantum number $k$.  This can be shown by recognizing that the quasimomenta $k$ describe the eigenvalues $e^{ika}$ of the discrete displacement operator.  The controls we have at our disposal are collective controls of the lattice.  The challenge is to control an ensemble of particles in parallel even though they each undergo a different quantum evolution, which mathematically is identical to the problem of robust control  \cite{Steffen03,Chow09b,Motzoi09,Rebentrost09,Chow10,Safaei09,Jirari09}.

For complex problems such as controlling an ensemble it is generally difficult to find control sequences analytically.  An alternative is to use techniques from optimal control theory, where the fields are found numerically by optimizing a cost function, usually a state or gate fidelity \cite{Rice00,KrotovBook,BrysonHo,Khaneja05}.  These numerical approaches are very powerful because the optimization algorithms they use are mostly platform independent and easily extendable to account for constraints such as those of ensemble control.  Optimal control theory has been applied to the problem of robust or ensemble control in many different contexts like NMR \cite{Khaneja06, Borneman10, Skinner11, Zhang11}, many-body entanglement \cite{Platzer10}, spin-chains \cite{Wang10}, and spin systems \cite{Said09}.

The remainder of this paper is a follows.  In section \ref{ch:lattice} we introduce the one-dimensional lattice model, its band structure and explore the interactions we utilize to control this system. In section \ref{ch:controltheory} we develop how this inhomogenous system can be treated within optimal control theory. The results of our numerical control treatment are shown in section \ref{ch:results}.

\section{Physics of the Optical Lattice \label{ch:lattice}}

\subsection{Model}

We consider a physical model of the optical lattice similar to the experimental apparatus described in Refs~\cite{Maneshi08,Maneshi05,Maneshi10}. The optical lattice potential is generated by two vertical counter-propagating lasers with an incidence angle of  $49.6^{\circ}$ and is loaded with $^{85}{\rm Rb}$ atoms from an optical molasses at 10 $\mu$K.  The Hamiltonian for an alkali atom interacting with an optical lattice potential in 1-D is given by 
\begin{equation}
   H_0 = \frac{p^2}{2m} + U_0 \sin^2 (k_L x),
   \label{OLHam}
\end{equation}  
where $m$ is the mass of the atom, $k_L$ wave number of the lattice laser light and $U_0$ is the lattice potential depth.  In experiments, the depth is typically $18$-$30 E_R$ though we consider a wider range for this numerical study.  Here $E_R = \hbar^2 k_L^2/2m$ is the recoil energy, i.e., the kinetic energy the atom gains by absorbing a lattice photon.  At these lattice depths for typical atomic densities the atoms are essentially non-interacting \cite{Maneshi08} and thus we can simplify the model by considering the evolution of a single atom.

The dimensionless form of this Hamiltonian is given by,
\begin{equation}
	H_0 = p^2 + r \sin^2{ x } = p^2 + \frac{r}{2}(1-\cos{2x}),
	\label{dimenham}
\end{equation}
where $r=U_0/E_R$ is the lattice potential depth in units of recoil energies and $p$ is now the momentum in units of $\hbar k_L$.  Bloch's theorem states that a periodic Hamiltonian such as Eq.~\eqref{dimenham} has eigenstates of the form, $\ket{\psi_{n}^{(k)}}$  or  
\begin{equation}
\braket{x}{\psi_{n}^{(k)}} = \psi_{n}^{(k)}(x)e^{ikx},
\end{equation}
 where $n$ indicates the energy band and $k$ the quasimomentum and $\psi_{n}^{(k)}(x+ m \pi ) =  \psi_{n}^{(k)}(x)$ for all integers $m$.  These Bloch functions satisfy the Schr\"odinger equation 
 \begin{equation}
E^{(k)}_n \psi_n^{(k)} = H^{(k)}_0  \psi_n^{(k)}  
 \end{equation}
 with the now $k$-dependent Hamiltonian
\begin{equation}
H^{(k)}_0 = (p-k)^2 + \frac{r}{2}(1-\cos{2x}).  \label{eq:Hk}
\end{equation}
The resulting wavefunctions can be calculated using the central matrix method or the Bloch method \cite{ashcroftmermin}.  Of primary importance is that the quasimomentum is a good quantum number. The single-particle energy eigenstates of the optical lattice thus, from a control perspective, form an inhomogenous ensemble of discrete quantum systems with different parameters $k$ in the Hamiltonian of Eq.~\eqref{eq:Hk}. As we shall see later, the controls also conserve $k$, thus preserving this ensemble decomposition.

To control this system we introduce the laser parameters $\eta(t)$ and $\phi(t)$ which represent the ratio of intensity with respect to $r$ and the phase, respectively, of the lasers. Rewriting \eqref{eq:Hk} in terms of these new variables we obtain,
\begin{equation}
\begin{split}
H^{(k)}(t) =& (p-k)^2 + \frac{r}{2} \left[1+\eta(t) \right] \left[1 - \cos \left(2x + \phi(t) \right) \right]   \\
=&H^{(k)}_0 -\frac{r}{2} \left[ 1 - \cos(2x) \right]\\
&+  \frac{r}{2} \left[1+\eta(t) \right] \left[1 - \cos \left(2x + \phi(t) \right) \right]\\
=&H^{(k)}_0+ \frac{r}{2} \left[1 - (1 +\eta(t)) \cos \phi(t) \right] \cos(2x)\\
 &+   \frac{r}{2} (1 +\eta(t)) \sin \phi(t) \sin(2x) + \frac{r}{2}\eta(t).  
\end{split}
\end{equation}
In order to express the Hamiltonian
in the standard form for quantum control,
\cite{Khaneja05}, we reparameterize the control fields in terms of
\begin{equation}
\begin{split}
\alpha(t) =& \frac{r}{4} \bigg[ 1- [1 + \eta(t)] \cos{\phi(t)}  \bigg], \\
\beta(t) =& \frac{r}{4} \bigg[ [1 + \eta(t)] \sin{\phi(t)} \bigg] .
	\end{split}
\end{equation}
so that the total control Hamiltonian (neglecting the global phase) is
\begin{equation}
H^{(k)}(t) = H^{(k)}_0 +  2 \alpha(t)   \cos(2x)+   2 \beta(t)  \sin(2x).\label{eq:controlhamil}
\end{equation}
The lattice will be filled with multiple atoms occupying a range of $k$-values. Measurement of the success of a quantum operation can be performed in the manner suggested in \cite{Myrskog05}, which involves averaging single particle measurements over the entire ensemble.  Demanding that we find $\alpha(t)$ and $\beta(t)$ that perform the desired gate irrespective of $k$ ensures that any gates we find will retain their fidelity after this averaging.

\subsection{Dispersion}

\begin{figure}[b!]
\centering
\includegraphics[width=1.0\columnwidth]{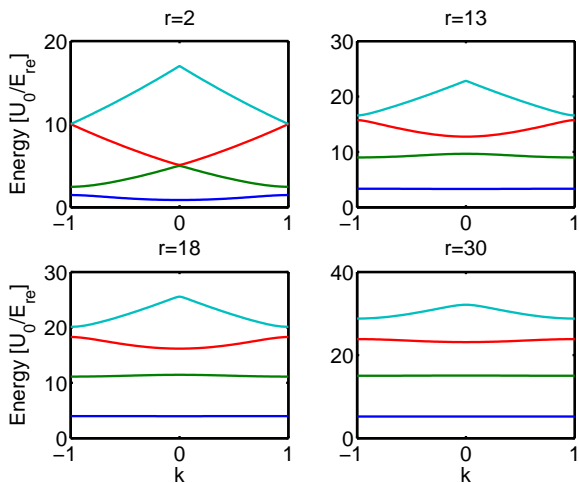}
\caption
{(color online) The first four energy 1-D band structures are shown in the momentum basis for four different potential depths $r$. The lower potential depth $r=2$ (top left) shows the largest amount of energy dispersion with the closest energy crossings. The depth $r=30$ (bottom right) shows a lesser amount of dispersion and larger energy splittings.}
\label{fig:energy}
\end{figure}

States with different quasimomenta evolve independently under the applied Hamiltonian $\hat{H}_0^{(k)}$ and so that they can be viewed as an ensemble of non-interacting systems, labelled by the quasimomentum.  The eigen-energies of this system depend on the quasimomentum, i.e.~$H_0^{(k)} \ket{\psi_{n}^{(k)} } = E_{n}^{(k)} \ket{\psi_{n}^{(k)} }$, leading to dispersion of the resonance frequencies across different quasimomenta.
 That is the resonant frequencies of excitations associated with different vertical transitions are not the same, see Fig.~\ref{fig:energy}.  We characterize the dispersion by the dimensionless quantity 
 \begin{equation}
	\mathcal{D}=1-\frac{\Delta E_{01}^{(1)}}  {\Delta E_{01}^{(0)}},
\end{equation}
where $\Delta E_{01}^{(k)} = E_1^{(k)}-E_0^{(k)}$.  In addition to energy dispersion there is dispersion in the strengths of the couplings between the bands with respect to the quasimomentum.
 
 The main consequence of the dispersion is that a single resonant harmonic excitation cannot simultaneously resonantly excite all transitions. This poses a formidable challenge for controlling an ensemble with high dispersion. For the optical lattice, the amount of energy dispersion decreases with an increase in the strength of potential, Fig. \ref{fig:energy}.  Unfortunately, the anharmonicity of the lattice, which is essential to resolve different transition frequencies, also decreases with increasing potential depth.  This leads to a trade-off when choosing the depth of the lattice between dispersion and anharmonicity.  Note that the latter alone can be addressed with techniques like DRAG \cite{Motzoi09}.

\subsection{Relation to the Charge Qubit}

A device that behaves similarly to the optical lattice in the context of superconducting qubits \cite{You05b,Insight} is the charge qubit \cite{Nakamura99, Bylander11, Koch07,Houck09}.  The effective Hamiltonian of this system takes the form
\begin{equation}
    \hat{H}^{(n_g)} = 4E_C(\hat{n} - n_g)^2 - E_J \cos{\hat{\phi}}
\end{equation}
where $\hat{n}$ is the number of Cooper pairs on the superconducting islands, $\hat{\phi}$ is the phase difference to the island, $E_C$ and $E_J$ are the charging and Josephson energy, respectively and $n_g$ is the effective offset charge.

Since $\hat{n}$ and $\hat{\phi}$ are canonically conjugate variables, we can directly map this Hamiltonian onto that of the optical lattice by letting
\begin{equation}
\hat{n} \rightarrow p, \quad \hat{\phi} \rightarrow x \quad \textrm{and} \quad \frac{E_J}{E_C} \rightarrow \frac{r}{2} .
\end{equation}
The only remaining component is to let the gate charge, $n_g$, play the role of the quasimomentum \cite{Schon90}.  The gate charge takes on a single value, unlike the the quasimomentum, but has uncertainty due to noise from fluctuating charges and voltages \cite{Houck09}. These fluctuations are slow and in particular they occur on time scales comparable to the duration of an experiment. An ensemble in these experiments, which are carried out on single quantum systems, is built up by repeating the experiment. In order to control the charge qubit we must account for this inherent spread over $n_g$.  The solution to this problem for the transmon qubit is to work in a lower dispersion regime, with a tradeoff in the effective anharmonicity\cite{Koch07,Houck09}.  The optimization techniques for these systems will be identical to the optical lattice, even though the causes of the dispersion are very different: ensemble averaging versus parameter uncertainty.  Note that in superconducting qubits, the CORPSE pulse sequence\cite{Cummins00} has been applied to mitigate fluctuations of the energy splittings \cite{Collin04}, 
which has been investigated further numerically \cite{Mottonen06}.

\section{Technique for Optimization\label{ch:controltheory}}

In order to find pulses that prepare a desired state or unitary transfer we use numerical techniques from optimal control \cite{Khaneja05,Brumer03,Machnes2010,Chakrabarti07,Rice00}.  Primarily we use a variant of the GRAPE (gradient-ascent pulse engineering) algorithm to numerically find optimal control fields for a fixed duration pulse \cite{Khaneja05}.  GRAPE is a numerical optimization method that maximizes the fidelity, $\Phi$, with respect to the control fields, $\mbf{u}$, for generating gates or states from some initial configuration of the system.  For the majority of this section we will focus on gate preparation since it is of greater interest  and because the state preparation protocol is a straightforward modification.

The optimization follows the form of a gradient search, where at each iteration, $j$, the control fields (initialized to a value $\mbf{u}^{(0)}$ ) are updated according to 
\begin{equation}
\mbf{u}^{(j+1)} = \mbf{u}^{(j) } + \epsilon \nabla_u \Phi,
\end{equation}
where $\epsilon$ is a small step parameter, usually chosen adaptively for faster convergence.  An essential feature of an optimal control algorithm such as GRAPE is to compute the gradient $\nabla_u \Phi$ in an efficient manner \cite{Khaneja05,Machnes2010,Motzoi11}.  This protocol is not guaranteed to find a global optimum; however for many problems it is easy to find high-fidelity solutions after optimizing a small number of initial guesses \cite{Fouquieres11,Machnes2010}.

We require a suitable fidelity function to average the Hamiltonian over the ensemble. As a starting point we take the standard average gate fidelity, 
\begin{equation}
\Phi = \frac{1}{d^2}|\Tr (\tilde{V}^\dagger U)|^2\label{eq:grapephi}
\end{equation}
where $\tilde{V}$ is the desired gate that we wish to implement, $U$ is the total time-evolution due to our choice of control fields and $d$ is the dimension of the Hilbert space.  Our desired gate should act on the $\ket{\psi_n^{(k)}}$ states independent of $k$, so we can write it as $\tilde{V} =  \int \ud k V \Pi(k)$, where $\Pi(k)$ is the projector onto the quasimomenta in the first Brillouin zone and $V$ is an operator that acts only on the energy band degree of freedom, thus commuting with $\Pi(k)$.   If we define 
\begin{equation}
U^{(k)} = \mathcal{T}\left[ e^{-i \int_0^T \ud t H^{(k)}(t)} \right] 
\end{equation} 
then the total evolution operator of our system is $U = \int \ud k' U^{(k')} \Pi(k')$, leading to the fidelity
\begin{equation}
\begin{split}
\Phi =& \frac{1}{d^2}\left| \Tr \left(\iint \ud k \ud k^\prime \,  V^\dagger \Pi(k)  U^{(k')} \Pi(k') \right)\right|^2\\
=& \frac{1}{d^2} \left|\int \ud k \Tr \left(  V^\dagger  U^{(k)}\right)\right|^2,\\
\end{split}
\label{eq:intfidelity}
\end{equation} 
where the final trace is taken over only the band degree of freedom.

This fidelity is not the same as simply averaging the standard gate fidelity over the ensemble. An important feature of Eq.~\eqref{eq:intfidelity} is that the absolute value is taken after the ensemble average, forcing the phases for different members of the ensemble to be equivalent in order to maximize the fidelity.  This global phase on each $U^{(k)}$ at first seems like it should be irrelevant, but it plays an important role when it comes to optimizing pulses with a finite bandwidth.  A simple example where this global phase is important is the following:  restrict to evolutions in $SU(d)$ where all Hamiltonians are traceless and all unitary evolutions therefore have $\det (U) = 1$.  In this setting there is a finite set of unitary maps that maximize $|\Tr (V^\dagger U)|$ namely $U = e^{i 2 \pi q/d} V$, where $q=1\ldots d$ and we assume $\det (V)=1$.  These are the only global phases for which $U$ has a unit determinant.  Ignoring this phase and attempting to optimize a gate that is invariant to a parameter such as the quasimomentum can lead to a situation where the global phase locks to different values in different regions of $k$-space.  The finite bandwidth of our controls does not allow $U^{(k)}$ to change instantaneously with $k$, and so there will be some intermediate regimes between two values of the global phase where the gate performs very poorly.  The evolution of the optical lattice we have outlined in this paper (Eq.~\ref{eq:controlhamil}) is not restricted to $SU(d)$, though we suspect that a similar argument can be made. This requires further investigation.  In practice our numerical optimization can get trapped if we neglect this global phase.  We see that there are high fidelity regions in $k$ that are connected by intermediate regions with low fidelity.  Our fidelity accounts for this issue by insisting the global phase to be uniform across the ensemble.

In order to make this fidelity tractable for numerical evaluation we sample the quasimomentum for a discrete number of values ${k_1,k_2, \ldots k_M}$ and then we arrive at the fidelity 
\begin{equation}
\Phi= \frac{1}{(dM)^2} \left|\sum_{l=1}^M  \Tr \left(  V^\dagger  U^{(k_l)}  \right)\right|^2.
\label{eq:gatefidelity}
\end{equation}
The Nyquist sampling rate sets a minimum bandwidth of the control pulses to ensure a minimum fidelity with respect to variations in the quasimomenta.  Therefore if we sample the quasimomentum on a fine enough grid we can be assured that the intermediate values have a high fidelity, and as verification we can test this hypothesis after optimizing our fields. 

For GRAPE we must provide the derivative, $\partial \Phi / \partial u_q$, which we calculate by
\begin{equation}
\begin{split}
\frac{\partial \Phi}{\partial u_q}=& \frac{2}{(d M)^2} \textrm{Re} \left( \sum^{M}_{l=1} \Tr \left[ V^\dagger \frac{\partial U^{(k_l)}}{\partial u_q} \right] \right. \\
&\left. \times \sum^{M}_{l'=1} \Tr \left[ V^\dagger U^{(k_{l'})} \right]^* \right).
\end{split}
\end{equation}
Essentially, we must calculate $ \partial U^{(k_l)} / \partial u_q $ for $M$ different evolutions leading to a computational-scaling for evaluating this derivative that is only $M$ times more than standard GRAPE on a $d$-dimensional system.  However, this is significantly faster than optimizing a gate for a $d M$-dimensional Hilbert space since multiplying $d \times d$ matrices scales as $\mathcal{O}(d^3)$.

\section{Results\label{ch:results}}
\begin{figure}[htbp]
  \begin{center}
    \subfigure[]{\includegraphics[width=1.0\columnwidth]{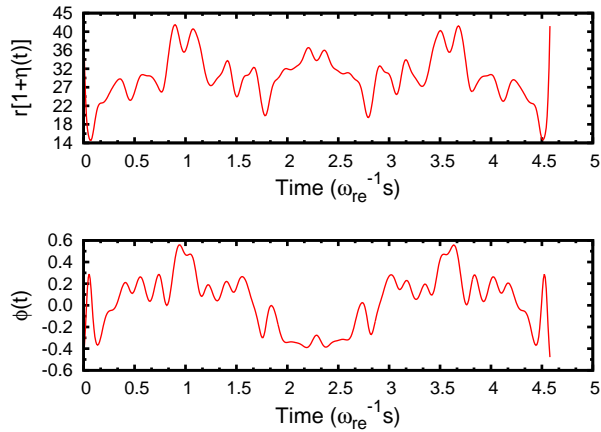}\label{fig:controlsa}}
    \subfigure[]{
    \includegraphics[width=1.0\columnwidth]{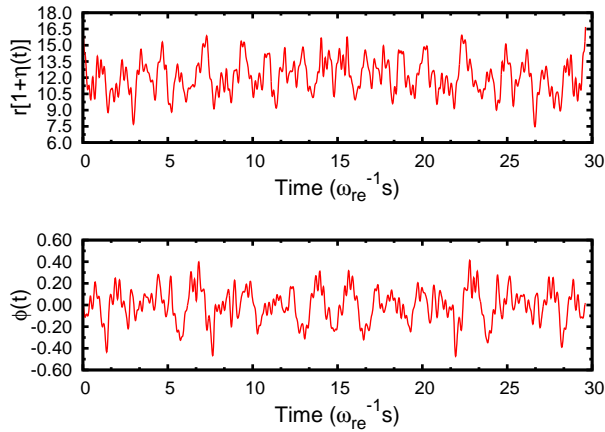}\label{fig:controlsb}}
\caption[Evolution of a two state system with positive coupling using a palindromic pulse]
{(color online) Comparison of two optimized controls for preparing a $X_{\pi}$-gate when the optimization uses only ten points in quasi momentum space.  Both pulses are well-behaved in the sense that they don't translate the lattice by a full lattice site or possess very large excursions from the initial value for $r$.  (a) An optimized pulse for a potential depth $r=17$ with 5.4\% dispersion and fidelity 99.3\% which is calculated by sampling over 100 quasi momentum values.  The duration of the pulse was 5 free oscillations (at k=0). (b) An optimized pulse for potential depth $r=12$ and dispersion 13.2\% with a sampled fidelity of only 66.8\% .  The time duration of the pulse was 25 free oscillations (at k=0).}
\label{fig:controls}
  \end{center}
\end{figure}

\begin{figure}[htbp]
  \centering
 \includegraphics[width=1.0\columnwidth]{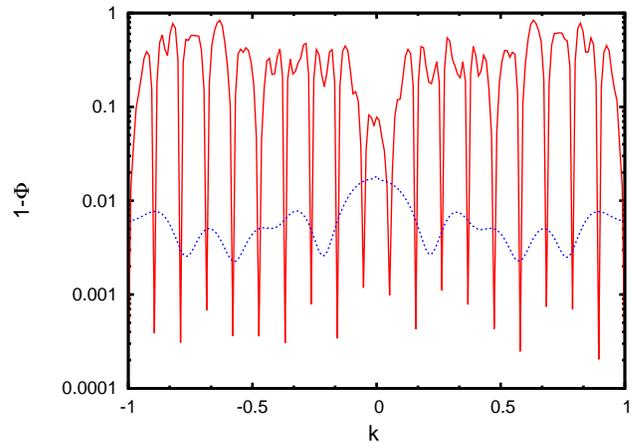}
    \caption{(color online) Comparison of gate error for well-behaved optimized controls in Fig.~\ref{fig:controls} where the blue dotted line corresponds to control sequence (a) and the red solid line to (b). For (a) the gate error is at most 1.8\% across the possible quasimomenta and the average gate error is 0.6\%.  In (b) the gate error has been minimized for the quasimomenta points that were sampled for optimization but was 66.8\% on average when sampled over 100 points. }
    \label{fig:controlfidelities}
\end{figure}

Using the techniques outlined in the previous section, we optimize control parameters for the optical lattice Hamiltonian in Eq. (\ref{dimenham}) for varying potential depths and pulse durations.  Our target gate is an $X_{\pi}$-gate on the first two energy bands.  We consider both a Rabi pulse assuming no dispersion and bounded random controls fields as the initial guesses for the control fields.  The evolution is simulated using only the first six energy bands from our model and sampling over 20 values of the quasimomenta.  While we sample over only a small number of values for the quasimomenta in order to have efficiently performed the numerical search, when we calculated the final fidelity we sampled over a finer grid in quasimomentum space. 


Here are two examples of optimized controls for performing $X_{\pi}$-gates on our optical lattice model.  One is a control for an optical lattice system with potential depth $r=17$ (which corresponds to a dispersion of 5.4\%), as shown in Fig. \ref{fig:controlsa}.  The duration of this control was 5 free oscillations (at k=0).  The gate error across the ensemble was less than 1\% with the exception of particles with quasimomentum near zero, Fig. \ref{fig:controlfidelities}.  Even with such high dispersions in the energies and control Hamiltonians we are able to find gates with reasonable fidelities.   

The second control in Fig. \ref{fig:controlsb} is for an optical lattice system with potential depth $r=12$ and thus a dispersion of 13.2\%.  The average fidelity for the optimized points was 99.96\% but after finer examination of the fidelity across the quasimomenta it was found to have an average of 66.8\%, as shown in Fig. \ref{fig:controlfidelities}.  Thus a more cautious approach should be taken with regards to optimizing a coarse sampling over the quasimomenta space.  These limitations can be overcome by finer sampling over the range of quasimomenta, however, this can become computationally expensive.

\begin{figure}[htbp]
\centering
\includegraphics[width=1.0\columnwidth]{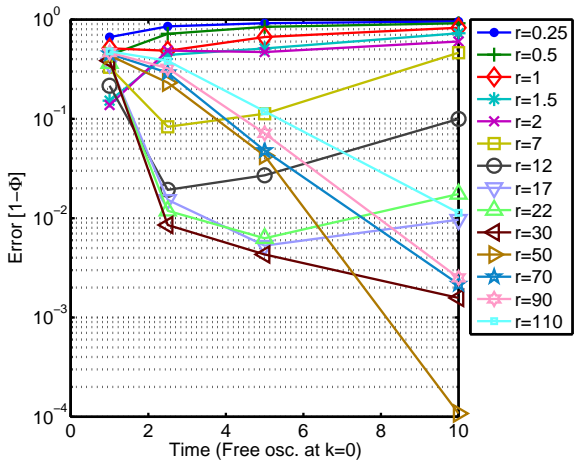}
\caption
{(color online) The maximum fidelity for an optimized pulse over a range of times, from 1 to 10 free oscillations (at k=0), preparing a $X_{\pi}$-gate as a series of different potential depths.  The potential depth ranged from $r=0.25$ to $r=110$, giving a dispersion ranging from 93.8\% to 0.01\%, respectively.  Each point is an average of optimized pulses from 1 Rabi and 10 random initial pulses. }
\label{fig:gatefidelityshort}
\end{figure}

We observe the relationship between the maximum fidelity of the solutions and the level of dispersion of the Hamiltonian.  Ideally, our gradient search will halt when a local maximum in the fidelity is reached but this may take a considerable amount of computational resources.  All our optimizations were halted after either the algorithm converged to a solution or $10^5$ updates in the control parameters were performed.  Nevertheless, for short times (Fig.~\ref{fig:gatefidelityshort}) we find excellent control fields ($\Phi > 0.90$) for a range of potential depths from $12 \leq r \leq 110$.

\begin{figure}[htbp]
\centering
\includegraphics[width=1.0\columnwidth]{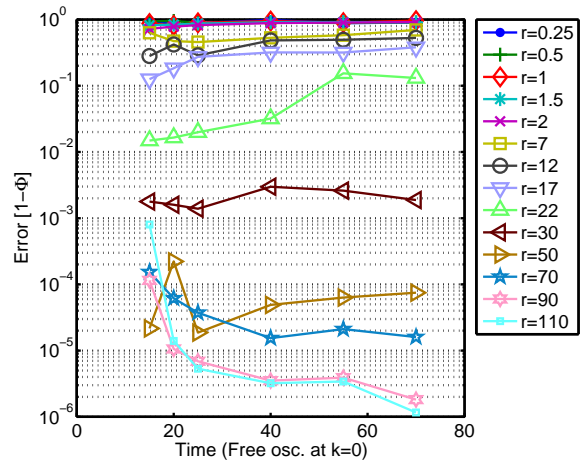}
\caption
{(color online) Same as Fig.~\ref{fig:gatefidelityshort} but with times from 15 to 70 free oscillations (at k=0). }
\label{fig:gatefidelityvsr}
\end{figure}

 In general, for long gate times, the maximum fidelity solution for $X_{\pi}$-gates became lower as the dispersion was increased,  Fig. \ref{fig:gatefidelityvsr}.  The optimization becomes less tractable with higher dispersion since we are asking a single set of control fields to solve substantially different problems depending on where the system is in quasimomentum space.  As a result, broadband pulses must be tailored to accommodate a range of possible energies and couplings.
 
We also find that for high dispersion the fidelity becomes worse as the gate time becomes longer.  This may seem counterintuitive, but it is due to the fact that as the gate time increases the fidelity can vary more quickly as a function of $k$.  For very short gate times, the fidelity is fairly constant across quasimomentum space, but for long gates we find that the fidelity is only high for the specific points we optimized, and dips to almost zero in the intermediate regimes, as seen in Fig.~ \ref{fig:controlfidelities}.  

\begin{figure}[htbp]
\centering
\includegraphics[width=1.0\columnwidth]{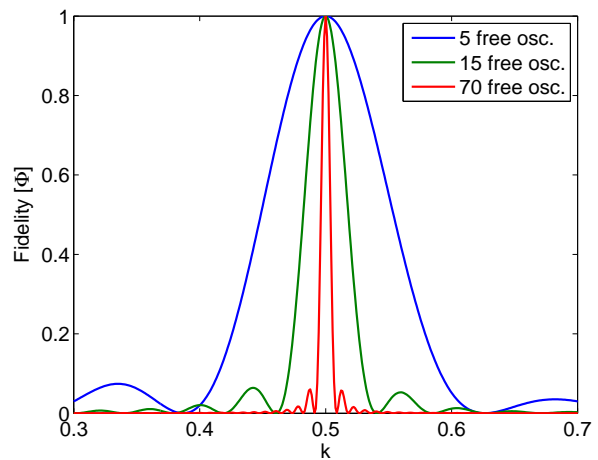}
\caption
{(color online) Pulses for performing $X_{\pi}$-gate at $r=2$ for three different times that were optimized specifically for k=0.5 and the fidelity response over quasimomenta shown.  This shows the effect of Nyquist bandwidth limits of the pulses on fidelity.  Shorter pulses have larger spectral bandwidth and thus affect a larger range of quasimomenta.}
\label{fig:nyquist}
\end{figure}

We can see the connection between gate duration and rate of change of fidelity very easily through the simulation described in Fig.~\ref{fig:nyquist}.  Here we run GRAPE for exactly one value of the quasimomentum and and then look at the performance of this control field across $k$.  We find that as the duration of the gate increases the fidelity function becomes more tightly peaked about the single value we have optimized.  In principle, there should alway be higher fidelity control fields at longer gate times but finding these fields becomes computationally expensive as we need to sample more values in quasimomentum space or simply run the algorithm from more initial conditions in the hope that one produces a flat fidelity curve.

\section{Conclusion}

We simulated our controls to observe the fidelity for all possible quasimomenta in the ensemble.  We also observed the minimum time to prepare a $X_{\pi}$-gate for a given dispersion using our techniques.  This shows that it is possible to perform operations of high fidelity in reasonable times for a system with inhomogeneity in both energies and couplings.  Specifically, we have demonstrated control of a shallow one-dimensional optical lattice with atoms that have inhomogeneous control matrix elements and distinct spectra.  With an ensemble averaged optimal control technique we have demonstrated robustness and shown that with 13.2\% dispersion in energies it is possible to achieve fidelities as high as 98.3\% with relatively short pulses.  In lower potential depths, some of the added dispersion can be corrected allowing to take advantage of greater anharmonicity to perform shorter gates.  

The technique does contain limitations due to coarse sampling of the ensemble for optimization.  With finer improvements, this should be resolved.  We have also shown that this technique performs better for shorter times with high dispersion due to the flatter response of the fidelity across the ensemble.  

Where developing control pulses from analytic techniques can become intractable for complicated systems, optimal control provides a way out.  The control techniques outlined in this paper are not exclusive to the optical lattice and can be applied to other quantum systems.  In fact, these techniques can be applied to any system with noise in control parameters, energy fluctuations, ensembles with varying couplings and energies, or of uncertain system specifications.  With further enhancement of numerical and analytical techniques, there is no reason that higher fidelities and shorter times should be impossible.  Much work has been done in this field however there is potential for improved techniques for designing robust control pulses, allowing for greater control over quantum systems \cite{JrShin11}.  What has been shown in this paper is the application of optimal control to robust pulse design.

\begin{acknowledgments}
We acknowledge Chao Zhuang, Brian Mischuck, and Bill Coish for valuable discussions. J.M.G. was supported by CIFAR and DARPA/MTO QuEST through a grant from AFOSR. B.K., F.M., S.T.M., and F.K.W. were supported by NSERC through the discovery grants and QuantumWorks. 
This research was also funded by the Office of the Director of National
Intelligence (ODNI), Intelligence Advanced Research Projects Activity
(IARPA), through the Army Research Office. All statements of fact, opinion
or conclusions contained herein are those of the authors and should not be
construed as representing the official views or policies of IARPA, the ODNI,
or the U.S. Government.  This research was made possible in part by Sharcnet.

\end{acknowledgments}


\end{document}